# Paradigms in Physics Education Research

Amy D. Robertson, Rachel E. Scherr, and Sarah B. McKagan
Seattle Pacific University, Seattle, WA, 98119-1997

**Abstract.** In this paper, we describe two paradigms in physics education research (PER): recurrence-oriented and case-oriented PER. We connect theory on research methodologies in the social sciences to interviews with physics education researchers and examples of published PER to articulate the specific assumptions of recurrence- and case-oriented PER. We show that the different assumptions made in these two paradigms bear out in specific aims and research designs taken up by physics education researchers. In particular, recurrence-oriented research seeks reproducible, representative patterns and relationships; human behavior is modeled as governed by lawful (albeit probabilistic) relationships. Case-oriented research, in contrast, seeks to refine and develop theory by linking that theory to cases; human action is assumed to be shaped by the meanings that participants make of their local environments. We briefly offer examples in which researchers take up both the recurrence- and case-oriented research paradigms.

**Keywords:** research paradigms, physics education research, research methodology

## I. Introduction

Discipline-based education research (DBER) in STEM (science, technology, engineering, and mathematics) disciplines is proliferating, as disciplinary experts in physics, biology, engineering, etc., start to pursue education-oriented research questions *within* their disciplines. DBER fields – such as physics education research, biology education research, engineering education research, etc. – uniquely blend the expertise of the discipline with the methods, assumptions, and vision of education research. And as these fields proliferate, they often draw on more and more wisdom and approaches from the education research community.

Building community understanding and appropriate evaluative criteria for publication, promotion, etc., within DBER fields presents a unique challenge, since many discipline-based education researchers were trained in STEM disciplines, each with their own sets of standards for good research. This paper makes an effort to begin to delineate and describe two research paradigms within one DBER field, physics education research (PER).

By many accounts, PER was launched by the observation that students in introductory physics courses often do not learn what they are told (Beichner 2009; Cummings 2011; Docktor and Mestre 2011). First-generation physics education researchers investigated students' conceptual understanding by analyzing responses to Piagetian-style clinical interviews, open-ended written questions, and systematically-developed multiple-choice conceptual inventories. A few academic generations later, interests and approaches in PER have proliferated to include topics as diverse as the role of gender and race in physics learning (Lorenzo, Crouch, and Mazur 2006; Pollock, Finkelstein, and Kost 2007; Brewe et al. 2010; Kost-Smith, Pollock, and Finkelstein 2010; Kreutzer and Boudreaux 2012; Rosa and Mensah 2016), the impact of mathematics understanding on conceptual learning in physics (Meltzer 2002; Thompson, Bucy, and Mountcastle 2006; Pollock, Thompson, and Mountcastle 2008; Christensen and Thompson 2012), and the role of gesture and posture in understanding students' physics learning (Close and Scherr 2012; Scherr 2008). Researchers' repertoire of tools has likewise broadened to include video records (Lising and Elby 2005; Goertzen, Scherr, and Elby 2010; Frank and Scherr 2012) and statistical analysis (Chasteen et al. 2012; Chini et al. 2012; Henderson, Dancy, and Niewiadomska-Bugaj 2012), in addition to continued use of clinical interviews (Henderson and Dancy 2008; Podolefsky, Perkins, and Adams 2010; Watkins et al. 2012), classroom observations (Kreutzer and Boudreaux 2012; Turpen and Finkelstein 2010; West et al. 2013), and written assessments (Spike and Finkelstein 2012; Kryjevskaia, Stetzer, and Heron 2013; Robertson and Shaffer 2013).

In this paper, we describe two *paradigms* in PER – recurrence-oriented PER and case-oriented PER – that bear out in some of the diverse aims and research designs in the field. Using connections between interviews with physics education researchers and literature on research methodologies in the social sciences, including education research, we argue that these two paradigms draw on different assumptions about the social world and about what counts as rigorous, real, and/or trustworthy, when it comes to research accounts. We show how these assumptions are tied to the specific aims and methods of recurrence-oriented and case-oriented research. In referring to these as paradigms, we adopt Greene and Caracelli's (1997) definition of a paradigm as a set of assumptions and beliefs about "knowledge, our social world, our ability to know that world, and our reasons for knowing it" (6) that frame and guide a particular orientation toward research, "including what questions to ask, what methods to use, what knowledge claims to strive for, and what defines high-quality work" (6). In particular, recurrence-oriented research seeks reproducible, representative patterns and relationships; human behavior is modeled as governed by lawful (albeit probabilistic) relationships. Case-oriented research, in contrast, seeks to refine and develop theory by linking that theory to cases; human action is assumed to be shaped by the meanings that participants make of their local environments. We chose not to call these paradigms "qualitative" and "quantitative" so as not to conflate them with *methods;* paradigms, as we think of them, are fundamentally about *assumptions*. These assumptions may – and often do – bear out in choices about methods, but "paradigm" and "methods" are not interchangeable.[1]

To be clear, we do not mean to claim that there are *only* two paradigms in PER. Instead, we mean to highlight that (at least some of) the diversity that we see in aims and methods is *motivated by* differences in assumptions about how the social world works, and/or what counts as trustworthy or real when it comes to making claims about social phenomena, including physics learning. In other words, our primary argument is about plausible links between aims, methods, and paradigms, and we use our depictions of two such paradigms in PER to illustrate it.

Further, we briefly illustrate that these paradigms are neither incompatible nor mutually exclusive; many of the researchers we interviewed have taken up *both* recurrence-oriented *and* case-oriented research, either sequentially or simultaneously, and often to study the same phenomenon. Some researchers describe this – the taking up of different paradigms – as foundational to how they think about research. We argue that doing so does not constitute a third paradigm, if we define paradigm in a way consistent with Greene and Caracelli (1997), in that we do not see it as instantiating a set of assumptions about the social world, etc., that are distinct from recurrence- or case-oriented PER.

This work has implications for multiple audiences. For physics education researchers, our primary claim that – in more general terms – research that looks different may be motivated by different assumptions is relevant to our engagement with and assessment of one another's work. Many physics education researchers were trained in physics, a community with an "extreme culture of objectivity" (169) in which the relative merit of ideas is often assessed through competition (Traweek 2009). For such physicists, case-oriented research, which values the subjective, lived experiences of research participants, and often makes claims on the basis of detailed analysis of single cases, may seem to be poor-quality research. Meanwhile, for physics education researchers who centralize the context-dependent, subjective experiences of participants, recurrence-oriented research that is seeking lawful, population-level relationships may also feel like poor-quality research, in that it can seem to ignore people's real experiences in those contexts. Understanding not just that there are many different *methods* in play in our field, but that choices around methods may be motivated by assumptions about *how the social world works* and *what is real and/or trustworthy in research accounts* may help us to not only understand one another better, but also to develop appropriate measures of evaluation for publication, promotion, and so on. This feels especially important in the current political climate in the United States (Hochschild 2016), where a significant community of physics education researchers live and do their work. In an era when national public discourse includes questions about the extent to which science is trustworthy, we believe the PER community, a community that draws on both disciplinary expertise in physics and research that transforms education, will be more effective and transformative if we capitalize on the diverse strengths of our community, to send a consistent message of clarity, coherence, and hope. We think one path toward this ideal begins with appreciating and understanding our intellectual diversity.

---

[1] To briefly elaborate, the space of research is often parsed out according to *methods*: the processes by which researchers obtain and analyze data. A paradigm, in contrast, is a set of assumptions (which may bear out in research aims and methods). The same method may serve different paradigms. For example, interviews may be used for recurrence- or case-oriented research; the former may involve searching for recurring patterns across interviews, whereas the latter may involve trying to understand the meaning a student is making in a single episode from a single interview.

For methodologists, this paper may offer a case of how some paradigms and assumptions within the broader education research community are being taken up and transformed by researchers who identify as physicists. Our primary approach to generating the claims in this paper, as we will describe in Section II, has been to illustrate connections between (1) physics education researchers' depictions of their own work in interviews and (2) methodological literature from the social sciences, including education research. And for education researchers who are seeking interdisciplinary collaborations with STEM discipline-based education researchers (but who do not themselves identify primarily with STEM), this paper may provide a useful starting place for cross-talk.

Sections III and IV unpack our primary claim: that the recurrence- and case-oriented research paradigms make different assumptions about the social world and about what counts as rigorous or real, and that these assumptions bear out in the specific aims and methods of recurrence-oriented and case-oriented research. Making this claim requires us to both (1) articulate the distinct assumptions associated with each paradigm and (2) plausibly link these assumptions to the specific aims and research designs taken up by physics education researchers, as in Figure 1. We recognize that the relationship between paradigm, aims, and research design is more complex than Figure 1 implies; aims may inform research design, and vice versa, and the arrows we have drawn can go the other way. Our point here is to highlight the relationship between paradigm and aims/design, not to comprehensively express the relationships between the entities in the figure.

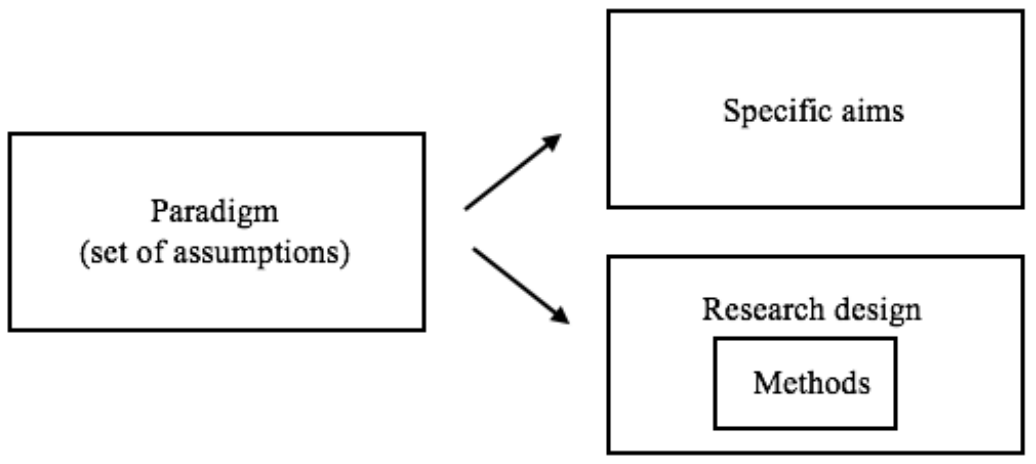

**Figure 1.** Structure of plausibility argument that paradigms can motivate specific aims and research designs. Here, methods are subsumed under research design; we discuss specific designs (e.g., experimental design, naturalistic observation), and these designs often include the use of specific research methods (e.g., video recording, surveys).

In Section V, we briefly illustrate ways in which particular physics education researchers take up both case- and recurrence-oriented research.

## II. Methods

Our primary effort in this paper is to describe two different paradigms in PER – that is, two different sets of assumptions about "knowledge, our social world, our ability to know that world, and our reasons for knowing it" (Greene and Caracelli 1997, 6) – and to argue that these different sets of assumptions plausibly motivate different choices of aims and methods within PER. We also show, briefly (in Section V), that these two paradigms are not incompatible or mutually exclusive by illustrating ways in which researchers draw on them both to understand physics teaching and learning.

These claims grow out of connections we made between (1) researchers' own descriptions of their work during interviews and (2) methodological literature in the social sciences, including education research. In a sense, our claim is not new; methodologists already know that research assumptions motivate research choices. On the other hand, our work takes a careful look at what this looks like in one field – PER – including how it is expressed in researcher talk. This field may present a particularly informative case in that the researchers in question identify not just as education researchers but also as members of the physics community, which has its own cultural norms tied to what counts as research (Harding 1991; Traweek 2009). In this section, we describe our methods in detail.

*Motivation*. This work was motivated by the first author's transition from large-*N* pattern-seeking research (as a graduate student) to case study analysis (as a junior faculty member). A cursory read of the methodological literature [*e.g.,* Creswell (2009), Otero and Harlow (2009)] offered her steps and processes, but she wanted to understand how to think like a case study analyst – *e.g.,* to learn what makes a good case, what purposes this kind of work may serve, and so on. She began to interview physics education researchers, and the insight that these interviews offered her evolved into the substance of this paper.

*Interview sampling*. Early on, we identified researchers that we thought of as conducting either quantitative or qualitative PER, and our choice of interview subjects was aimed at helping us flesh out what each of these meant. What began to emerge from the interviews was that interviewees' depictions of their work and motivations did not entirely overlap with what is traditionally thought of as quantitative or qualitative research, especially if conceived in terms of data and methods. Thus, we chose to foreground research paradigms in our analysis, and to call the two paradigms we describe "recurrence-" and "case-oriented," to highlight what we considered to be a central focus of each paradigm: the priority of recurrence or the importance of cases. As we began to develop our characterizations of the recurrence-oriented and case-oriented paradigms, we noticed that some of the research that our interviewees depicted, as well as other examples of prominent PER, did not precisely fit either description. Subsequently, we conducted additional interviews with individuals whose research we perceived as not fitting into either the recurrence-oriented or case-oriented paradigms. The illustrations in Section V – of research that combines paradigms – come from these latter interviews.

In total, the first author conducted eighteen interviews with physics education researchers. Interview participants were chosen on the basis of their perceived research interests, faculty status, relationship to the first author, and willingness to be interviewed. Because an original goal of this project was to better understand what we now call case-oriented research, researchers that we perceived to be conducting this kind of research made up a larger fraction of the interview subjects.

*Content of interviews*. Each interview lasted between forty-five minutes and one hour and was conducted either in person, by remote video, or on the phone. The interviews were loosely scripted. Major topics included: the kinds of questions each interviewee is interested in answering, the process by which each interviewee tries to answer these questions, the kinds of claims each interviewee seeks to make, what counts as evidence for these claims, and the criteria each uses to evaluate his or her research. Each interview was recorded, content-logged, and summarized (Derry et al. 2010). The summaries were sent to individual interviewees and revised on the basis of their feedback.

*Interpretive framework and analysis of interviews*. The claims in this paper grew out of interactions between the content of our interviews and literature on research methodologies in the social sciences (including education research). Some of the assertions we make throughout the paper are descriptive in nature: they attempt to characterize aims and methods, as described by our interviewees. Such assertions are lower-level inferences: themes that were pulled directly from transcripts and summaries of interviews. In each case, the illustrative aims and research designs we use reflect the examples that came up in our interviews, and are not meant to be a complete list of aims or methods in recurrence- or case-oriented PER.

Our claims about research *assumptions,* on the other hand – what they are, which moments in which interviews embody them, and how they are tied to and/or motivate some of the aims and methods of recurrence- and case-oriented research – are higher inference. These were generated out of connections between methodological literature and the content of our interviews. Each one – the literature and the interviews – helped us to understand the other. In writing this paper, we take specific perspectives or assumptions expressed within the methodological literature – e.g., that case studies are meant to broaden audience perspective (Wehlage 1981; Donmoyer 1990; Maxwell 1992), or that the recurrence of a result across independent observers lends plausibility to the truth of the result (Cook 2002) – and illustrate what these assumptions look like in PER, and how they are plausibly tied to particular aims and methods in the field. Further, we *foreground* those assumptions from the methodological literature that were evidenced by our interviews; these are the assumptions that comprise the recurrence- and case-oriented paradigms we describe.[2] In short,

[2] To be clear, our claims do not comprise a literature review; we have not taken a single paradigm from the education research literature (e.g., constructivism, post-positivism) and mapped it onto PER. Our data suggests that this would not work, at least not in a one-to-one mapping kind of way. We suspect this is in part because PER is historically rooted in physics, which translates into the foregrounding of particular facets of education research paradigms. It is also likely a limitation of our interview sampling, and of the

the literature helped us to articulate the assumptions within each paradigm, and the interviews helped us to demarcate which aspects of the methodological literature were relevant and/or applicable to research within PER, as described by our interviewees.

We identify this manuscript as case-oriented research. We show how (methodological) theory manifests in the concrete details of researcher talk about their own work (Erickson 1986), and we make a theoretical claim that research assumptions are tied to choices of method and aim within one field (PER). Further, our effort as we conducted our interviews and analysis was to understand how the meaning that researchers make of what they are doing shapes the substance of their research, drawing on the assumptions of case-oriented research we articulate in Section IV.

In the remainder of the paper, we use examples of published physics or science education research to illustrate our characterizations of recurrence-oriented research, case-oriented research, and ways of combining these paradigms. Examples were selected because they clearly embody either the recurrence-oriented or case-oriented research paradigm or a way of combining them; appear in a journal recognized by the PER community as a primary site for publishing research; and are authored by physics education researchers whose work is recognized as shaping community standards. These papers were selected after our analysis of interviews; they were chosen to (and did) validate and extend our original claims.

Prior to submission of our manuscript, we conducted extensive member checks (Maxwell 2004; Otero and Harlow 2009; Creswell 2009), offering interview participants and authors of the published examples we use opportunities to provide us feedback on drafts of this paper, and we revised the manuscript on the basis of their feedback. Interviewees resonated with the content of our descriptions of the paradigms, although some objected to being labeled as committed to a single paradigm, or to having their research reduced to a single label. As such, and to indicate that researchers may simultaneously or sequentially participate in different research paradigms, we focus on *research* commitments, rather than *researcher* commitments throughout the paper. For example, a researcher may believe that social action is shaped by the meaning that students are making of their local environment (assumption within the case-oriented paradigm), while also believing that across many students, that action (or those behaviors) can be understood in terms of probabilistic patterns or relationships (assumption within the recurrence-oriented paradigm).[3] However, in our interviews, some researchers expressed strong personal commitments to the premises, values, and assumptions reflected in our characterizations of a specific paradigm, suggesting that it is possible for researchers to primarily identify with one research paradigm at a given time.

*Limitations*. As with most case-oriented research, our claims draw on small *N* – in this case, interviews with a small number of physics education researchers – which limits their generalizability to the *population* of physics education researchers *writ large*. However, we have been clear that we are not aiming to characterize PER comprehensively; we do not think our work does this. We mean to *illustrate* how certain assumptions (organized into paradigms) show up within PER and may plausibly link to choices of aim and method. These are theoretical claims that may have broad applicability, but they do not require large numbers to substantiate.

Further, we acknowledge (in fact, appreciate) that our work is *one* perspective about how to characterize or describe research within PER. As above, our characterization is not comprehensive, and it is likely oversimplified – we have focused on a particular aspect of research that was depicted within our interviews.

---

nature of our data (i.e., researchers' reflections on their work, in response to particular questions, over the course of one hour).

[3] We acknowledge an anonymous reviewer's help in shaping the location and substance of this sentence.

Relatedly, the content of our interviews was shaped by the first author’s interest in understanding research paradigms. When research assumptions – or ways in to research assumptions – came up in the natural course of the interviews, the first author focused on and followed up on this. Likewise, in her content logging and attempts to understand interviewees’ points-of-view, this was a (sometimes unconscious, sometimes more explicit) focus of her interest and attention. This narrows the scope of what researchers discussed during the interviews, and thus what we can infer, in a broad sense, about their work, from these interviews.

# III. Recurrence-oriented physics education research

Recurrence-oriented research premises include (1) that human behavior is guided by predictable relationships between variables and (2) that real phenomena are reproducible. Thus, recurrence-oriented research locates generalities in recurring patterns and relationships by looking for trends in aggregate data (Figure 2). Recurrence-oriented *physics education* research instantiates these premises by identifying recurring teaching and learning phenomena and instructional causes and effects, often in the service of helping instructors plan and predict instruction and/or specific learning outcomes. Researchers often do so using large-scale surveys and controlled experiments.

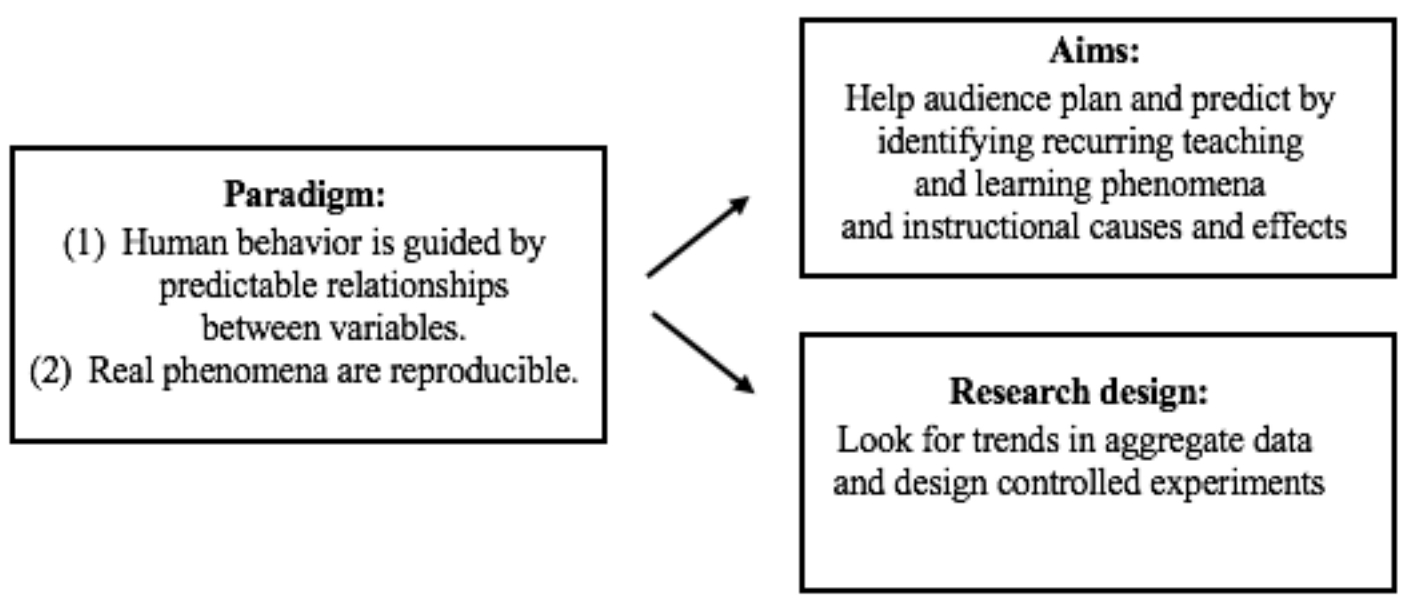

**Figure 2.** Connections between recurrence-oriented research paradigm, aims, and design.

Interviews with physics education researchers and published research by Pollock and Finkelstein (2008) inform and illustrate our characterization of recurrence-oriented PER. In “Sustaining educational reforms in introductory physics,” Pollock and Finkelstein build on an earlier study (Finkelstein and Pollock 2005) demonstrating that *Tutorials in Introductory Physics* (McDermott, Shaffer, and the PEG at the University of Washington 2011), a PER-based curriculum, could be successfully implemented in a context other than the one in which it was developed. In the manuscript we discuss in this paper (Pollock and Finkelstein 2008), the authors explore intra-institutional hand-off of *Tutorials*. They demonstrate that (1) student learning gains [as measured by <*g*>, average normalized gain (Hake 1998)] are consistently large when *Tutorials* are used in recitation sections at the University of Colorado at Boulder, and that (2) there is some variation in <*g*> that may be attributable to curricular choice and/or faculty familiarity with PER.

## A. Paradigm: Research is predicated on the assumptions that (1) human behavior is guided by predictable relationships between variables and that (2) real phenomena are reproducible.

In this section, we will explore some of the specific assumptions about “knowledge, our social world, our ability to know that world, and our reasons for knowing it” (Greene and Caracelli 1997, 6) that are often tied to the choice to locate generalities in recurring patterns and relationships, as in recurrence-oriented PER.

### 1. Human behavior is guided by predictable relationships between variables.

Recurrence-oriented research often models human behavior as governed by predictable relationships between variables. Such relationships are understood to be most accurately reflected at the level of populations, which consist of all members of the group of interest. Since obtaining data from an entire population is usually not possible, relationships are apprehended using probabilistic and statistical tools (Erickson 1986; Guba and Lincoln 2005) that allow researchers to evaluate the possibility of making incorrect population-level inferences on the basis of data collected from a sample (Roth 2009; Willis 2007; Firestone 1993). This kind of research is modeled after that in the natural sciences, in which nature's uniformity allows a mechanical, chemical, or biological understanding of causation (Erickson 1986; Donmoyer 1990; Mishler 1979; Willis 2007; Schofield 1990). This uniformity does not imply linearity; rather, it implies that variation in human behavior follows a trend (Field 2009; Ding and Liu 2012). Thus, any non-uniformities will likely average out to zero if one considers an entire population (or a representative sample of that population).

In recurrence-oriented PER, both the human and physical worlds are modeled as governed by lawful relationships, with the caveats that (1) there are many more variables to consider in social interactions than in physical ones and (2) the interactions between these variables are much more complex. One researcher described his research as follows:

> "I'm trying to figure out the underlying dynamics [of physics learning and teaching], and yes, I think the basic approach is similar to the standard physics research approach, which is to try to understand the various factors involved in the system, to try to control a reasonable number of them and to vary certain others to look at the outcomes, with an aim to understanding the underlying dynamics."

Pollock and Finkelstein (2008) likewise search for a relationship between variables to explain semester-by-semester differences in introductory physics students' average normalized gain. For example, they highlight the effect of the variable 'instructor' on <*g*>, binning instructors according to their familiarity with PER. The authors relate "faculty background" to student learning, writing,

> "…we observe that in instances when PER faculty are involved in instruction, in either the lead or secondary role, students post the highest learning gains. When PER-informed faculty…are involved in instruction, students post higher learning gains than when only traditional faculty are involved." (6)

### 2. Real phenomena are reproducible.

The probabilistic and statistical tools used in recurrence-oriented research embed specific assumptions about what it means for a claim to be true. In particular, recurrence-oriented research represents human behavior in terms of observable phenomena and predictable relationships that exist "independent of [the scientists'] personal values and sociopolitical beliefs" (Moss et al. 2009, 502). To ensure that observations and inferences truly reflect these phenomena and relationships – and not biases that result from an unrepresentative sample or from the personal values of the researcher – researchers conducting recurrence-oriented PER privilege phenomena that recur over and over, independent of observer and context (Schofield 1990). Cook (2002) ties *recurrence* to *truth,* saying that even though "observations are never theory-neutral, many of them have stubbornly re-occurred whatever the researcher's predilections" and thus have "such a high degree of facticity that they can be confidently treated as though they were true" (179). One of our interviewees stated:

> "In general, I say, 'That's a very interesting result. Now do it again and see what happens.' And if you get it a third time and if it's similar to what you observed the first two times, then you can begin

> to believe that you're onto something. But if you do it a second and a third time and it is very different than what happened the first time,…then you have to be very skeptical and say that there's a good chance that this was just a random fluctuation type of thing."

One of the central questions of Pollock and Finkelstein's (2008) study concerns the reproducibility of large conceptual gains when *Tutorials* are implemented across and within institutions. The first two graphs in the paper show the statistical indistinguishability of results from (1) implementations of *Tutorials* at (a) the University of Washington (*Tutorials* development site) and (b) CU-Boulder (*Tutorials* implementation site) and (2) the (i) first and (ii) second implementations of *Tutorials* at CU-Boulder. These graphs communicate that the gains achieved (1) at UW and (2) during the first implementation at CU-Boulder do not represent random fluctuations or irreproducible, extenuating circumstances; they represent a *real* curricular effect.

## B. Aims: Research helps readers plan and predict by identifying recurring teaching and learning phenomena and instructional causes and effects.

Tied to the assumptions described above, recurrence-oriented PER frames understanding and shaping physics learning in terms of predictable patterns and relationships: what can instructors expect, what variables affect learning, and how might we manipulate these variables to shape particular instructional events? The overarching aim is to help the audience plan and predict by identifying recurring teaching and learning phenomena. For example, one researcher we interviewed said that:

> "it's important to sort of map out these possible ways that students think about these pretty basic ideas...because…if an expert teacher had sort of a better sense of the…ways that students might think about these topics and the kinds of errors that they might make, that it makes for better teachers…If a teacher can hear a response and…have it not come out of the blue, have thought about, 'Okay, this, this is something that's been described before.'"

Our interviewees described specific research directions that these aims take. For example, these researchers:

- *Identify the conceptual difficulties that students may encounter when learning topic X* (*e.g.*, treating current as used up by bulbs in a circuit (McDermott and Shaffer 1992)). Though some phases of this research explore the flavor of student ideas, the aim is often to determine and describe the patterns and parameters that characterize a population (Ding and Liu 2012; Pyrczak 2006). Some interviewees explicitly describe the purpose of such research as helping instructors to anticipate the ideas that students may struggle with.

- *Assess the effectiveness of instructional materials* in improving student performance on conceptual questions. For example, one interviewee describes the goal of her research as "provid[ing] materials that are going to, on average, have a positive effect for some kind of average group of students."

- *Determine which variables influence: (1) learning gains and (2) misconception-like patterns in student responses*. Pollock and Finkelstein's (2008) manuscript illustrates this purpose when it poses "factors" – including "faculty background and the particular curricula used in recitation sections" (7) – that "contribut[e]" (7) to variation in student learning gains. The authors indicate their commitment to measuring *average* effects when they justify their use of <*g*> as an assessment of student learning:

  > "…[measurements of average gains] allow us to compare, *in aggregate,* the impact of different course implementations by different faculty on the overall performance of students enrolled in these courses." (3, emphasis added)

The literature affirms all three of these purposes for recurrence-oriented research, clarifying that such research may be “descriptive (assigning numbers or category labels to data on particular variables) or relational (investigating the relationship between two or more variables in the sample)” (Maxwell and Loomis 2003, 254).

## C. Design: Researchers look for trends in aggregate data and design controlled experiments.

To ensure that observations and claims *truly* reflect general phenomena and population-level relationships – rather than personal bias or a skewed sample – recurrence-oriented research privileges recurring, reproducible phenomena and relationships. For example, researchers conducting recurrence-oriented PER interpret patterns in aggregate student responses to written and interview questions as indicating specific conceptual difficulties with particular physics topics. The generalizability of these patterns is substantiated by asking the question in multiple contexts:

> “…To try to ensure that what is coming out is not…*only* an artifact of the question, and that if you never asked the question that way, it would never…come up,…you have to ask a variety of different questions. And even if you get similar responses to different questions, it doesn't mean that you've uncovered some, like, robust [knowledge structure]...But you've got something that is more than just…a reaction to a particular question. Like there's something there. It may not be completely robust and coherent, but it's going to come up under a variety of circumstances.”

Connected to the assumption that human behavior is guided by predictable relationships between variables, recurrence-oriented research often seeks to understand the dynamics of learning by reproducing experiments that test cause-and effect relationships. For example, our interviewees described several experiments they used to demonstrate that variable $x$ (*e.g.,* instructional intervention) influences outcome $y$ (*e.g.,* performance on conceptual exam, misconception-like reasoning, etc.). One researcher described how he determines which variables affect certain misconception-like patterns in student responses. In order to ensure that these patterns do not reflect a misinterpretation of a question, he asks questions in multiple contexts and eliminates those that produce idiosyncratic results. When these questions have been eliminated, and knowing already what kinds of patterns emerge in response to the remaining questions, he and his colleagues tweak the questions to see what happens to the response patterns.

In designing survey questions and experiments, sampling is often intentional. Researchers choose groups that are representative of a particular population (*e.g.,* introductory physics students or K-12 teachers) so that they can infer population-level relationships and descriptive parameters from their sample-level measurements (Pyrczak 2006).

To be clear, our description of recurrence-oriented research design is oversimplified, emphasizing the specific relationship between research paradigm and design. In practice, recurrence is not the only criterion that researchers apply in choosing to report (or trust) a result. For instance, researchers also seek plausible mechanisms that link cause and effect variables, and the patterns they report are often those that are tied to fundamental physics ideas (and are thus instructionally significant). As an example of the former, Pollock and Finkelstein (2008) propose that the relationship between faculty background and successful *Tutorials* implementation may be the product of alignment between curricular goals and adaptation practices:

> “In order to implement a course practice, one must attend to specific details of the local environment, institutional specifics that tend to vary with time, and make a myriad of decisions that are not prescribed or even documented – instructors must adapt their approaches and associated curricular practices. For faculty members well versed in the field of physics education research, and familiar with the development of the specific innovations, in this case tutorials, this adaptation can happen in a manner that is informed and aligned with the curricular goals.” (6)

Further, single informal observations may inform later stages of research, in which the researcher investigates whether a phenomenon is recurring or reproducible.

## IV. Case-oriented physics education research

Premises of case-oriented research include that social actions are guided by the meanings that people are making of their local environments and that reality is subjectively constructed. Individuals conducting case-oriented research thus immerse themselves in the details of a local context, separating the universal from the particular by connecting case to theory. Case-oriented PER specifically aims to use cases to illustrate, build, and/or refine theory, generating theoretical claims by naturalistically observing local interactions as they take place (Figure 3).

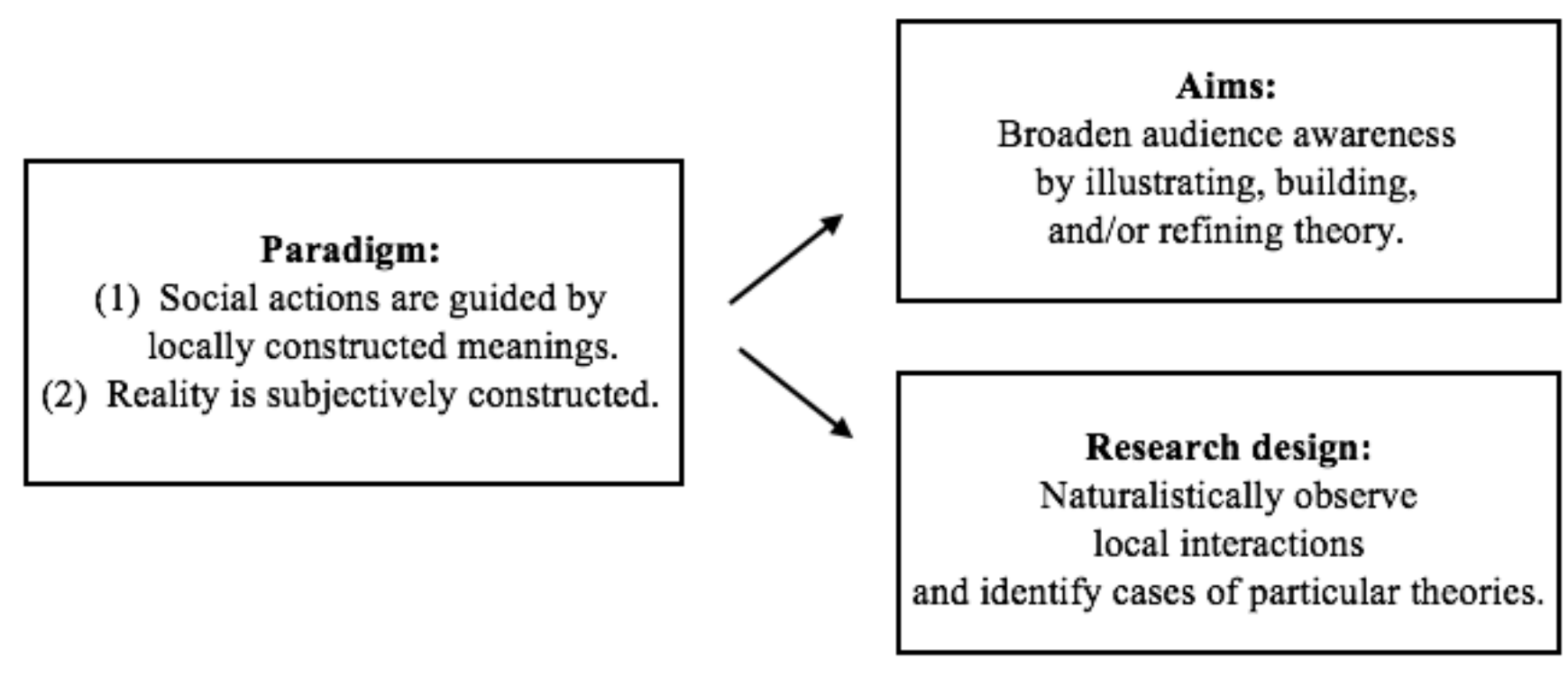

**Figure 3.** Connections between case-oriented research paradigm, aims, and design.

Interviews with physics education researchers and an example of published science education research inform and illustrate our characterizations of case-oriented PER. Berland and Hammer's (2012) case study, "Framing for Scientific Argumentation," presents three episodes from a sixth-grade science class. The authors focus on the social and epistemic expectations constructed by the students and teacher. In the first episode, the "idea sharing discussion," Mr. S (the teacher) nominates students to contribute ideas, often acknowledging or revoicing these ideas without evaluating them. In the second "argumentative" discussion, which takes place later in the unit, students engage with one another's ideas and try "to persuade each other to accept their claims" (74). The "discordant" discussion takes place immediately after the "argumentative" one, when Mr. S "move[s] to resume his role as epistemic and social authority" (74). Unlike in the first two episodes, in the third, "instabilities" emerge from the competing expectations of students and teacher: some participants' expectations were more consistent with the framing of the idea-sharing discussion, whereas others' expectations were more consistent with the framing of the argumentative discussion. The authors connect student and teacher framings in these three episodes to the literature on argumentation and suggest that certain frames are better aligned with productive argumentation practices.

## A. Paradigm: Research is predicated on the assumptions that (1) social actions are guided by locally constructed meanings and that (2) reality is subjectively constructed.

Now we turn to specific premises that are often tied to the choice to generalize on the basis of a single (or a small number of) cases, as is often done in case-oriented PER.

### 1. Social actions are guided by locally constructed meanings

Case-oriented PER is guided by the assumption that people construct locally meaningful interpretations of their environments (Erickson 1986; Guba and Lincoln 2005; Creswell 2009; Willis 2007; Firestone 1987); that people take action on the basis of these interpretations (*i.e.,* these interpretations are causal) (Erickson 1986; Bredo 2006); and that culture can organize interactions and promote shared meanings among groups of people that regularly interact (Erickson 1986; Anderson-Levitt 2006). The meanings that participants make of their experiences are dynamic and exist only in the context of local interactions, evolving as they continually (1) make sense of (and shape) their contexts and (2) respond to other participants who are simultaneously making sense of (and shaping) the context (Erickson 2007; Mishler 1979). Case-oriented PER understands social phenomena in terms of "what people mean and intend by what they say and do" and how these meanings are connected to and shaped by "historical, cultural, institutional, and immediate situational contexts" (Moss et al. 2009, 502). Researchers choose events for analysis that highlight the social mediation of meaning and/or that reveal local patterns that organize interaction. For example, one interviewee described her ongoing analysis of interviews with middle school students:

> "There were many students who throughout the course of the interview would sometimes use chemistry vocabulary words. And then at other times, they would switch, where they would start drawing on their everyday experiences…And what I have been thinking about that is that that is an epistemological issue, so that actually, students aren't quite sure how to engage in these interviews that we use so regularly as researchers. And that they're sort of trying on a number of different ways that they could engage in the interview."

In this quote, the researcher attributes students' participation and knowledge-on-display to the meaning that they are making of their local context.

Berland and Hammer's (2012) analysis also embeds this orientation toward social action. The authors document take-ups and dismissals of ideas that are linked to the meaning students and teacher are making of their shared activity. For example, in the "argumentative" discussion, students "frequently addressed one another directly and responded to each other's arguments" (79) (rather than directing their contributions to their teacher, Mr. S). The authors infer stable epistemic and social expectations throughout the discussion: students expect to assess ideas on the basis of evidence and reasoning and to hold ideas in opposition to one another (so that one idea prevails). They also expect to select ideas for further discussion and to control turn-taking themselves. Thus, when Mr. S intervenes to quiet the influx of student contributions, they ignore him, and he acquiesces. His bid for a shift in social expectations – toward himself as moderator of the discussion – is not taken up *because* it is inconsistent with the meanings the students are making of the discussion.

### 2. Reality is subjectively constructed.

The assumption (discussed immediately above) that social action is guided by meanings that are locally constructed and dynamically evolving is tied to the assumption that reality is subjectively constructed – that these locally constructed meanings are themselves what is *real* to participants, and thus what matters for a research account. Thus, case-oriented research tends to *trust* accounts that attend to the details of context and highlight multiple layers of meaning. Researchers conducting case-oriented research tend to foreground interactional complexity:

> "There was just so much going on in [*Tutorials*] that I had been unaware of. My former thing about pre- and post-testing was just missing so much of the amazing stuff that was *really* happening in the tutorial…Ever since, my research has been organized around trying to appreciate the complexity of what is happening in a fascinating classroom."

Berland and Hammer's (2012) study highlights the importance of participants' social and epistemic expectations to the conversational dynamics and to the productivity of students' argumentation practices. In the authors' analysis, the relevant context in which the discussion takes place is the *meanings* that students and teachers are making of "what is going on" in their shared space. These meanings are participants' *real* experience of the context and thus what guide the unfolding dynamic of the discussion. And these meanings are complex – tied to social and epistemic expectations and to verbal and nonverbal signals by which participants communicate their expectations to one another.

## B. Aims: Research broadens audience perspective by illustrating, building, and/or challenging theory.

The assumptions described above influence the ways in which case-oriented research seeks to understand and shape physics teaching and learning. Researchers conducting this kind of research assume that in a particular instance of a teacher teaching, some aspects of what occurs are "absolutely generic, that is, they apply cross-culturally and across human history to all teaching situations" (Erickson 1986, 130). However, the *way* in which these generic aspects manifest themselves are intimately tied to the local context. The task of the researcher is to "uncover the different layers of universality and particularity that are confronted in the specific case at hand – what is broadly universal, what generalizes to other similar situations, what is unique to the given instance" (Erickson 1986, 130). By identifying what is universal about particular instances of teaching and learning – instances that are *cases* of specific theories – and using those instances to build, refine, and/or challenge theory, researchers conducting case-oriented PER intend to broaden reader perspective (Donmoyer 1990). The hope is that readers will not only become aware of new events that they might attend to, but also deepen their vision for the analytical complexity of events that might on the surface seem ordinary (Schoenfeld 1992). One researcher we interviewed said:

> "My goal is to say at the end of the day that I've expanded the perspective that a reader of my work might have. That at the beginning they wouldn't have thought that this [event] was something of interest, then notice this set of things that happened, the complexity of it, the richness of it, and notice how much we could be paying attention to or are paying attention to when we naturally interact with the world."

Interviewees who described case-oriented PER identified a number of ways in which they seek to refine, build, or challenge theory, including:

- *Demonstrating possibility* (Turner 2004; Eisenhart 2009). For example, one researcher said:

  > "When I see great things, *those* are the things I want to talk about, almost like existence proofs, like 'this is possible.' [One distinctive moment] shows that this sort of instruction is possible."

- *Clarifying participants' points of view* (Firestone 1993, 1987). Like an anthropologist characterizing a remote island culture, a researcher conducting case-oriented PER may see aspects of students' experience that are "so customary for [them] that they were held outside of [their] conscious awareness" (Erickson 1986, 123), such as assuming that learning physics consists of properly locating and manipulating equations (Hutchison and Hammer 2010; Scherr and Hammer 2009).

- *Revealing and challenging implicit assumptions* [or "conventional views" (Ragin, Nagel, and White 2004, 17)]. For example, one interviewee framed his study of the dynamics of learning as consistently challenging the assumption that "understanding" – of physics content, for example – is a stable state.
- *Developing mechanisms that explain why certain teaching and learning phenomena take place* (Firestone 1993). For example, one physics education researcher described his interest in the "mechanisms by which students reason," defining these as "possible invisible models…that explain individual's actions and behaviors."
- *Coordinating multiple modalities to better understand thinking and learning*. One researcher said that he wants "to figure out not just the verbal information but body language and gesture and tone of voice and all of these other things that convey meaning," in line with research in linguistics and embodied cognition (Tannen and Wallat 1993; Goodwin 2000; Lakoff and Nuñez 2000).

Berland and Hammer's (2012) case study serves many of these functions. The authors demonstrate that it is possible for young students to productively engage in scientific argumentation; develop mechanisms – in this case, framings – that explain the dynamics of three classroom discussions; and coordinate multiple modalities, including verbal and nonverbal signals, by which participants communicate their expectations to one another.

## C. Design: Researchers naturalistically observe local interactions and identify cases of particular theories.

The case-oriented research paradigm asserts that theoretical accounts are trustworthy when they are grounded in and clearly tied to the details of context and when meanings are made sense of in terms of local interactions. Thus, researchers conduct case-oriented research by naturalistically observing learning as it takes place; selecting cases that have theoretical significance; constructing a narrative that brings readers close to the context; and connecting case to theory by analyzing the event in light of its theoretical significance, articulating how it adds to, challenges, or refines theory (Maxwell 1992).

The specific questions and purposes (*e.g.,* challenging implicit assumptions or demonstrating possibility) that a case-oriented research analysis will serve are often not determined in advance but stem from the data itself (Ragin et al. 2004), and claims are usually an outcome rather than the beginning of the research process (Blee 2004). One researcher we interviewed articulated the beginning of her research sequence as follows:

> "The process starts when somebody with good researcher or teacher eyes sees something that wows them [and then goes on to] look for other things in the video that maybe seem similar to you, so that you can start to say, "Listen, I don't know what it is, but I feel like these things all go together...I think these are all about the same thing. *What thing are these all about?*" To articulate your sense of what matters about the episode." (emphasis added)

Research processes include data collection (often video or audio records of instructional settings), textual documentation of the data in the form of field notes or content logs, episode selection, detailed analysis of discourse or behavior, and refinement of claims.[4] Episodes need not be representative; those that are illustrative, extreme, deviant, or revelatory may be equally of value, depending on the claim that they are used to support (Erickson 1986; Yin 2003; Engle, Langer-Osuna, and de Royston 2014; Schofield 1990; Firestone 1993). One researcher summarized the analysis process as determining "what is true" (the claim, based on the evidence) and "what matters" (the theoretical significance – *i.e.,* the theory/theories that the episode is a *case of*) about the episode.

---

[4] Again, here, our description of the case-oriented research process is simplified to highlight the relationship between paradigm and research design.

This research process is reflected in Berland and Hammer's (2012) manuscript. The authors say that they chose the "argumentative" and "discordant" discussions because they significantly deviated from the 'norm' with respect to argumentation and stability, and they chose the "idea sharing" discussion because it represented the 'norm' for this class. They formulated preliminary claims about "what was taking place [in the three discussions] and why" (74) by applying the "theoretical construct of framing" (74). They refined their claims by connecting the discussions to one another, "progressively refin[ing their] hypotheses" (74) in light of their deepening understandings of each one.

# V. Combining case- and recurrence-oriented research

So far, we have described two research paradigms in PER – recurrence- and case-oriented research – and argued that different aims and methods within this field may be *motivated by* or *tied to* different assumptions about the social world and what counts as real or trustworthy. We now suggest that it is both possible for (1) a researcher to be primarily committed to a *paradigm,* or set of assumptions, such that their choices of methods and aims grow out of that commitment, *and* it is also possible for (2) a researcher to be primarily committed to an *aim* (or set of aims) or *method(s),* and choose among and between paradigms that best answer to those aims/methods.[5] And this is not binary: researchers can move between these and/or hold commitments that span them.

These two possibilities were highlighted for us in our interviews. Many of our early interviewees foregrounded commitments and assumptions that spanned their work, highlighting possibility (1) for us, and helping us to start to characterize the recurrence- and case-oriented paradigms. As we expanded our pool of interviewees, some emphasized a deep commitment to a topic or theme within their work, highlighting possibility (2) for us. Of course, what was foregrounded in any given interview does not comprehensively reflect a given researcher or their work; we do not mean to imply that here. But this distinction in emphasis within our interviews highlighted for us the incompleteness of our developing characterizations of case- and recurrence-oriented PER in depicting the work of our interviewees.

This section is meant to broaden our discussion about paradigms, briefly. Our point here is in part to do justice to our data and to the researchers who shared their thinking about their work with us; our characterization of our interviews would be incomplete without this section. But it is also to show that both (1) and (2) above are possible – that although, as one of our interviewees said, "[my colleague who conducts case-oriented research] talked about how she's going to do qualitative research no matter what because that's just want she likes to do and what interests her," it is also true that researchers may (as she continues) "start with the questions [they] want to answer and…chose the research method to match the question." Further, we wish to show that the *way* in which paradigms are taken up in (2) is not *always* straightforwardly pragmatic. In other words, the logic model may not be about "meet[ing] the practical demands of a particular inquiry" (Rocco et al. 2003, 596). Researchers may also take up paradigms *dialectically* (Greene and Caracelli 1997; Rocco et al. 2003), so as to purposefully create tensions between assumptions and methods that allow for a more complex – and in some cases more realistic – depiction of a phenomenon. In the paragraphs that follow, we give examples – from PER – of each: research that more pragmatically combines the two paradigms we have described, and research that does so more dialectically.

---

[5] There is some debate as to whether or not *all* choices of method or paradigm are driven by a research question or aim (*e.g.,* Firestone (1987), Willis (2007)). In one sense, this may be true, in that a given research process may begin with a question, such as, "What conceptual difficulties do students encounter when learning topic *X*?," and then researchers may take up a research design that answers this question. In another sense, however, possible questions and research designs in recurrence-oriented and case-oriented research may be narrowed by specific paradigmatic assumptions; researchers may ask questions that are answerable by factors and prioritize recurrence (recurrence-oriented research) or that foreground locally-constructed meanings and prioritize accounts with multiple layers of meaning (case-oriented research).

**Pragmatically combining recurrence- and case-oriented research.** A number of our interview participants described their research in ways consistent with the literature's depictions of "pragmatic" mixed methods research.[6] In this approach, the choice of paradigm and method should first and foremost "'work best' to meet the practical demands of a particular inquiry…and thereby help to answer the research question" (Rocco et al. 2003, 596). This is reflected in a comment by one researcher we interviewed, who described her research as centering on faculty use of PER-based instructional strategies. She said,

> "I think you do what works. Like I think that you ask your question and then you go at it in a very organic way to figure out how to answer that question, and you use whatever perspectives and tools and methods you can."

This pragmatic approach is also reflected in a pair of papers written by Henderson and colleagues (Henderson 2005; Henderson, Dancy, and Niewiadomska-Bugaj 2012). The authors situate both papers in the larger agenda of understanding physics instructors' adoption and use of PER-based instructional strategies. Henderson (2005), which we would characterize as case-oriented PER, is a case study that ties one instructor's decision-making about a research-based instructional strategy to theory about the innovation-decision process. Henderson et al. (2012), which we would characterize as recurrence-oriented PER, analyzes results of a national survey, given to a representative sample of physics faculty, in terms of when and why faculty choose to discontinue use of research-based strategies. Each paper frames the choice of research design and assumptions *in terms of* the research questions and goals. For example, Henderson (2005) writes that the goal of his case study was to "develop a detailed understanding of the educational change process of one physics instructor as he attempted to change his instructional practices" (779), connecting the goal of acquiring a "detailed understanding" to his choice of a case study research design. The later study (Henderson, Dancy, and Niewiadomska-Bugaj 2012) seeks to "understand the extent to which faculty have been engaged in learning or implementing" (2) PER-based instructional strategies. Here, the authors seek a broad sense of which PER-based strategies faculty use – a population-level claim – and so administer a survey to a representative sample of physics faculty, intentionally designing the survey to measure variables of interest.

---

[6] In this section, we draw on literature about mixed methods research in the social sciences (particularly education research) to supplement and clarify the ways in which the physics education researchers we interviewed combine case- and recurrence-oriented research. Though we would agree that these researchers are *mixing methods,* their approach is distinct from the research literature's characterization of mixed methods education research in at least one important way. Although both embody a commitment to the question and an openness to and use of multiple methods, mixed methods educational research is defined by its use of more than one method in a single study or paper. The researchers we interviewed use multiple methods and take up multiple paradigms over the course of multiple studies, often to investigate a single question, but not necessarily in a single paper.

**Dialectically combining case- and recurrence-oriented research.** Other researchers, however, described their research in ways more consistent with the literature's characterization of "dialectic" mixed methods research. Dialectic mixed methods researchers in the social sciences argue that intentionally blending research paradigms is productive because of the tensions it generates (Greene and Caracelli 1997; Rocco et al. 2003):

> "…different paradigms do indeed offer different, and sometimes contradictory and opposing, ideas and perspectives. In dialectical mixed methods inquiry, these differences are valued precisely for their potential – through the tension they invoke – to generate meaningfully better understandings." (Greene and Caracelli 2003, 97)

Implicit in this depiction is the sense that neither case- nor recurrence-oriented research assumptions, in isolation, are sufficient to make sense of social phenomena, seeking to flesh out confirmations and contradictions between data sources. Dialectic research thus combines methods that uncover different facets of social phenomena. For example, one researcher who said that his research is "built around these overarching themes…that our students have these …cultural resources that they bring to the physics class that we need to understand better" also said:

> "I feel like I've always kind of tried to [use both qualitative and quantitative data]. There was some cool stuff in the quantitative data, but I felt like the story was not really there, the whole story, and I wanted to flesh things out. So I always felt like the quantitative data is cool, but I really need to have a bunch of interviews. Because it gave me more of a complete story."

# VI. Conclusion

In this paper, we have described two research paradigms in PER: recurrence-oriented and case-oriented research. We have argued that each is characterized by distinct premises and that these premises bear out in the specific aims and research designs taken up in recurrence- and case-oriented PER. We have given examples of ways in which researchers combine these two paradigms, pragmatically and dialectically, in pursuit of particular research themes.

As such, this paper is about meaning-making. We situate this manuscript in the case-oriented research paradigm, since much of our work has comprised connecting the practices and perspectives articulated by active members of the PER field to one another and to the theory and practices of education research more broadly. Although we expect particular instantiations of case-oriented PER, recurrence-oriented PER, and combinations of the two to vary from researcher to researcher, we believe that these instantiations will point back to the more general aims and premises of the research paradigms and approaches we have described here.

Our claims grew out of connections we made between (1) interviews with physics education researchers and (2) literature on research methods and premises in social science research. We thus implicitly claim our understanding of PER as a discipline can be informed by our understanding of research in the social sciences. An alternative approach would have been to study overviews of *physics* research methodologies (Marder 2011) or ethnographic accounts of physics research groups (Graves 1992; Latour 1987; Traweek 1988) and to connect PER to this literature. Focusing on these connections may foreground additional facets of the research paradigms we have articulated, or they may highlight altogether different ways of conceptualizing PER.

Our work has the potential to inform understandings across and within fields. For education research methodologists, our work may serve to illustrate which facets of education research paradigms are being taken up and transformed by a field that draws heavily on expertise in and cultural norms from physics. For education researchers seeking interdisciplinary collaborations with STEM discipline-based education researchers, our work may be a productive starting place for dialogue, and may support researchers in identifying overlaps and divergences in assumptions about how the social world works and what counts as a rigorous or trustworthy research account. For physics education researchers, our hope is that our work contributes to a community dialogue that lends itself to deeper understanding and appreciation of one another and our research efforts.

## Acknowledgments

We are deeply indebted to each of the physics education researchers that shared their thoughts with the first author, including Leslie Atkins, Andrew Boudreaux, Melissa Dancy, Brian Frank, Ayush Gupta, David Hammer, Danielle Harlow, Charles Henderson, Paula Heron, Andrew Heckler, Stephen Kanim, Sarah (Sam) McKagan, David Meltzer, Sanjay Rebello, Rosemary Russ, Mel Sabella, Rachel Scherr, and Michael Wittmann. We thank Paula R. L. Heron, David E. Meltzer, Justin C. Robertson, Eleanor Sayre, Thomas M. Scaife, Stamatis Vokos, and Michael C. Wittmann for thoughtful feedback on earlier versions of this manuscript.